\documentclass[sigconf,screen]{acmart}
\usepackage{graphicx}
\usepackage{adjustbox,lipsum}
\usepackage{multirow}
\usepackage{color,soul} 
\usepackage{cleveref}
\usepackage{todonotes}
\usepackage{url}
\usepackage{xspace}
\usepackage{enumitem}
\newcommand{\ie}{i.e.,\xspace}
\newcommand{\eg}{e.g.,\xspace}

\newcommand{\etal}{et al.\xspace}

\newboolean{showcomments}
\setboolean{showcomments}{true} 
\ifthenelse{\boolean{showcomments}}
{\newcommand{\nb}[2]{
\fcolorbox{black}{yellow}{\bfseries\sffamily\scriptsize#1}
{\sf\small$\blacktriangleright$\textit{#2}$\blacktriangleleft$}
 }
 
}
{\newcommand{\nb}[2]{}
 
}

\AtBeginDocument{%
\providecommand\BibTeX{{%
\normalfont B\kern-0.5em{\scshape i\kern-0.25em b}\kern-0.8em\TeX}}}

\setcopyright{acmcopyright}
\acmPrice{15.00}
\acmDOI{10.1145/3416505.3423564}
\acmYear{2020}
\copyrightyear{2020}
\acmSubmissionID{fsews20maltesquemain-p9-p}
\acmISBN{978-1-4503-8124-6/20/11}
\acmConference[MaLTeSQuE '20]{Proceedings of the 4th ACM SIGSOFT International Workshop on Machine Learning Techniques for Software Quality Evaluation}{November 13, 2020}{Virtual, USA}
\acmBooktitle{Proceedings of the 4th ACM SIGSOFT International Workshop on Machine Learning Techniques for Software Quality Evaluation (MaLTeSQuE '20), November 13, 2020, Virtual, USA}

\begin{document}

\title[DeepIaC: Deep Learning-based Linguistic Anti-Pattern Detection for IaC]{DeepIaC: Deep Learning-Based Linguistic Anti-pattern Detection in IaC}

\author{Nemania Borovits}
\email{n.borovits@uvt.nl}
\affiliation{
\institution{JADS, Tilburg University}
\country{The Netherlands}
}

\author{Indika Kumara}
\email{i.p.k.weerasingha.dewage@tue.nl}
\affiliation{
\institution{JADS, Eindhoven University of Technology}
\country{The Netherlands}
}

\author{Parvathy Krishnan}
\email{p.k.krishnakumari@tue.nl}
\affiliation{
\institution{JADS, Eindhoven University of Technology}
\country{The Netherlands}
}

\author{Stefano Dalla Palma}
\email{s.dalla.palma@uvt.nl}
\affiliation{
\institution{JADS, Tilburg University}
\country{The Netherlands}
}

\author{Dario Di Nucci}
\email{d.dinucci@uvt.nl}
\affiliation{
\institution{JADS, Tilburg University}
\country{The Netherlands}
}

\author{Fabio Palomba}
\email{fpalomba@unisa.it}
\affiliation{
\institution{University of Salerno}
\country{Italy}
}

\author{Damian A. Tamburri}
\email{d.a.tamburri@tue.nl}
\affiliation{
\institution{JADS, Eindhoven University of Technology}
\country{The Netherlands}
}

\author{Willem-Jan van den Heuvel}
\email{w.j.a.m.vdnHeuvel@uvt.nl}
\affiliation{
\institution{JADS, Tilburg University}
\country{The Netherlands}
}

\renewcommand{\shortauthors}{Borovits, N., \etal}

\begin{abstract}
Linguistic anti-patterns are recurring poor practices concerning inconsistencies among the naming, documentation, and implementation of an entity. They impede readability, understandability, and maintainability of source code. This paper attempts to detect linguistic anti-patterns in infrastructure as code (IaC) scripts used to provision and manage computing environments. In particular, we consider inconsistencies between the logic/body of IaC code units and their names. To this end, we propose a novel automated approach that employs word embeddings and deep learning techniques. We build and use the abstract syntax tree of IaC code units to create their code embedments. Our experiments with a dataset systematically extracted from open source repositories show that our approach yields an accuracy between $0.785$ and $0.915$ in detecting inconsistencies.
\end{abstract}

\ccsdesc[500]{Software and its engineering~Maintaining software}
\ccsdesc[500]{Computer systems organization~Cloud computing}
\ccsdesc[500]{Computing methodologies~Supervised learning by classification}

\keywords{Infrastructure Code, IaC, Linguistic Anti-patterns, Deep Learning, Word2Vec, Code Embedding, Defects}

\maketitle

\section{Introduction}

With growing importance for the ``need for speed'' in the current IT market, the software development cycle is becoming shorter every day.
Development and IT operation teams are increasingly cooperating as DevOps teams, relying massively on automation at both development and operations levels. 
The software code driving such automation is collectively known as Infrastructure-as-Code (IaC), a model for
provisioning and managing a computing environment using the explicit definition of the desired state of the environment in source code and applying software engineering principles, methodologies, and tools~\cite{morris2016infrastructure}. 

Although IaC is a relatively new research area, it attracted an ever-increasing number of scientific works in recent years~\cite{rahman2018gaps}. 
Nevertheless, most research has been done on IaC frameworks, while only a few studies explored the notion of infrastructure code quality. 
Among others, the first steps in this direction focused on applying the well-known concept of Software Defect Prediction~\cite{hall2011systematic} to infrastructure code defining defect prediction models to identify pieces of infrastructure that may be defect-prone and need more inspection.
In this perspective, previous works mainly focused on the identification of structural code properties that correlate with defective infrastructure code scripts.

However, this is only one of the possible proxies to identify defective code. 
Indeed, many problems can be rose by analyzing the plain text of software code.
In particular, linguistic anti-patterns, that is, recurring poor practices concerning inconsistencies among the naming, documentation, and implementation of an entity, have shown to be a good proxy for defect prediction~\cite{Linguistic, SPOT, DeepBug}.
Therefore, while the existing literature mainly focuses on structural characteristics of defective IaC scripts, none exists that analyze linguistic issues to the best of our knowledge. This motivation led to the research goal of this work: 

\begin{center}
\textit{How accurately can we detect linguistic anti-patterns in infrastructure as code (IaC) using a Deep-Learning approach?}
\end{center}

Boosted by the emerging trend of deep learning and word embeddings for software code analysis and defect prediction, we propose \textsc{DeepIaC}, a novel approach to detect linguistic anti-patterns in IaC, focusing on name-body inconsistencies in IaC code units.
Our experiments on a dataset composed of open source repositories show \textsc{DeepIaC} yields an accuracy between $0.785$ and $0.915$ in detecting inconsistencies with AUC (Area Under the ROC Curve) metric between $0.779$ and $0.914$, and MCC (Matthews correlation coefficient) metric between $0.570$ and $0.830$.
We deem our approach can contribute to step the current research up by tackling IaC Defect Prediction by a different perspective and providing a solid baseline for future studies focusing on linguistic issues.

\paragraph{Structure of the paper}
\Cref{sec:background} describes background and related works.
\Cref{sec:approach} details our approach to identify linguistic anti-patterns.
\Cref{sec:evaluation} elaborates on the empirical evaluation of the proposed approach, its results, and limitations. 
Finally, \Cref{sec:conclusion} concludes the paper and outlines future works.

\section{Background and Related Work}\label{sec:background}
This section introduces Infrastructure-as-Code and the Ansible configuration management technology and describes previous studies to identify defects and anti-patterns in infrastructure code. 

\begin{figure}[ht]
\centering
\includegraphics[width=0.6\columnwidth]{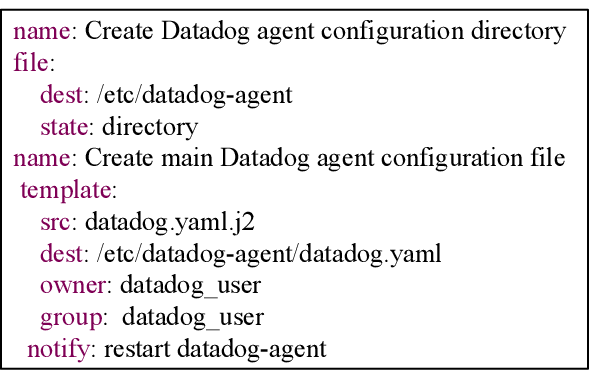}
\caption{A snippet of an Ansible role, showing two tasks}
\label{fig:iac_snippet}
\end{figure}

\subsection{Infrastructure-as-Code and Ansible}\label{sec:iac-ansible}
Infrastructure-as-Code (IaC) is a model for provisioning and managing a computing environment using the explicit definition of the desired state of the environment in source code and applying software engineering principles, methodologies, and tools via a Domain Specific Language (DSL). IaC DSLs enables defining the environment state as a software program, and IaC tools enable managing the environment based on such programs. In this study, we consider the Ansible IaC language, one of the most popular languages amongst practitioners, according to our previous survey~\cite{guerriero19adoption}.

In Ansible, a \textit{playbook} defines an IT infrastructure automation workflow as a set of ordered \textit{tasks} over one or more \textit{inventories} consisting of managed infrastructure nodes.
A \textit{module} represents a unit of code that a task invokes and serves a specific purpose, such as setting up a Datadog agent, creating a MySQL database, or installing an Apache webserver.
A \textit{role} can be used to group a cohesive set of tasks and resources that together accomplish a specific goal, such as installing and configuring MySQL.
When the tasks are executed, the states of the resources in the target nodes change.
To react to such changes, \textit{handlers} can be configured per task using \textit{notify} parameter.

\Cref{fig:iac_snippet} shows an Ansible snippet for configuring a Datadog agent.
The two tasks use the Ansible modules \textit{file} and \textit{template} to create a directory to keep the configuration file of Datadog and generate a configuration file from a template. Once the configuration file is created (\ie a state change), the handler is triggered to ensure that the Datadog agent is restarted to make the new configuration effective.

\subsection{Related Work}
Most of the previous works describe infrastructure code quality in terms of smelliness~\cite{folwer1999refactoring} and defects-proneness of Chef and Puppet infrastructure components.
From a smelliness perspective, Schwarz \etal~\cite{schwarz2018code}, Spinellis~\etal~\cite{spinellis-smells}, and Rahman \etal~\cite{rahman2019seven} applied the well-know concept to IaC, and identified code smells that can be grouped into four groups: (i) \textit{Implementation Configuration} such as complex expressions and deprecated statements; (ii) \textit{Design Configuration} such as broken hierarchies and duplicate blocks; (iii) \textit{Security Smells} such as admin by default and hard-coded secrets; (iv) \textit{General Smells} such as long resources and too many attributes. In \cite{WIMSDefect}, we developed an ontology-based approach to detect smells in TOSCA infrastructure as code.
From a defect prediction perspective, Rahman \etal~\cite{RAHMAN2019148} identified ten source code measures that significantly correlate with defective infrastructure as code scripts such as properties to execute bash and/or batch commands, to manage file permissions, and more.

In this work, we step up this research line by proposing a novel automated approach that employs code embeddings (vector representation of IaC code) and deep learning techniques to detect linguistic anti-patterns, focusing on name-body inconsistencies in IaC code units. 
We focus on Ansible, rather than Puppet and Chef, because Ansible is the most used IaC in the industry~\cite{guerriero19adoption}.

\begin{figure*}
\centering
\includegraphics[width=1\textwidth]{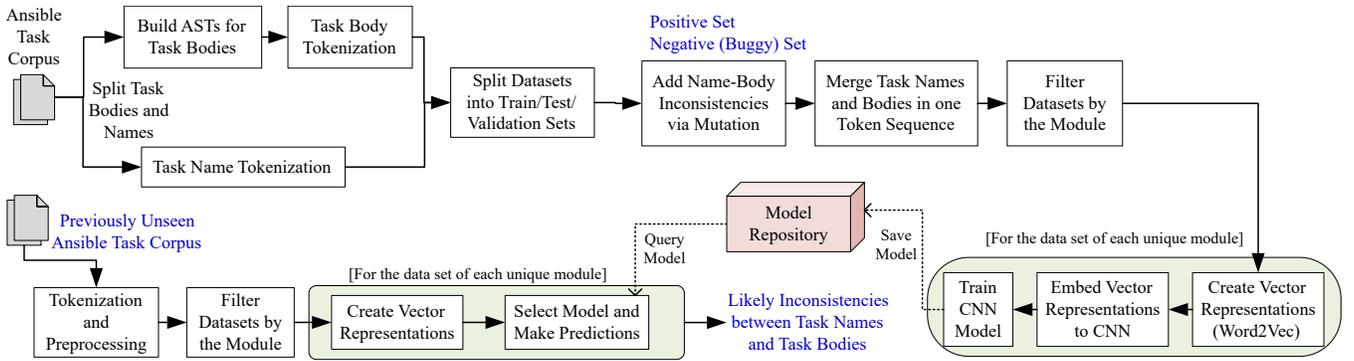}
\caption{Overview of the DeepIaC approach}
\label{fig:deepiac}
\end{figure*}

\section{Deep-Learning-based Linguistic Anti-Pattern Detection for IaC}
\label{sec:approach}

This section presents \textsc{DeepIaC}, our approach to identifying inconsistencies between names and logic/bodies in IaC code units and, in particular, in Ansible.
\Cref{fig:deepiac} illustrates the workflow of DeepIaC as a set of steps, which can be categorized into the following phases: 

\begin{description}[leftmargin=*]
\item[Corpus Tokenization] Given a corpus of Ansible tasks, this phase generates token streams for both task names and bodies. To tokenize a task's body while considering its semantic properties, we build and use its abstract syntax tree (AST). 

\item[Data Sets Generation] Since it is challenging to find a sufficient number of real buggy task examples that suffer from inconsistencies, we apply simple code transformations to generate buggy examples from likely correct examples. We perform such transformations on the tokenized data set and assume that most corpus tasks do not have inconsistencies. Indeed, several previous studies~\cite{DeepBug, 9054826SANER} in software defect prediction have successfully applied similar techniques to generate training and test data.

\item[From Datasets to Vectors] We employ \textsc{Word2Vec}~\cite{mikolov2013distributed} to convert the token sequences into distributed vector representations (code embeddings). We train a deep learning model for each Ansible module type as our experiments showed a single model does not perform well, potentially due to low token granularity. Thus, the tokenized data set is divided into subsets per module, and the code embeddings for each subset are separately generated. 

\item[Model Training] This phase feeds the code embeddings to a Convolutional Neural Network (CNN) model~\cite{matsugu2003subject} and train the model to distinguish between the tasks having name-body inconsistencies from correct tasks. The trained model is stored in the model repository. 

\item[Inconsistency Identification] The trained models (classifiers) from the model repository are employed to predict whether the name and body of a previously unseen Ansible task are consistent or not. Each task is transformed into its corresponding vector representations, which can be consumed by a classifier. 
\end{description}

\begin{figure}
\centering
\includegraphics[width=1\columnwidth]{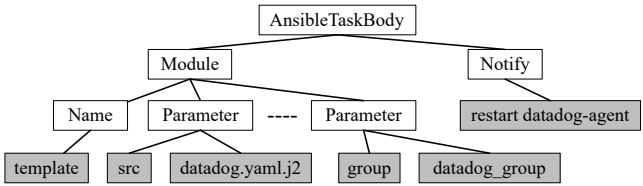}
\caption{AST model for a task using \textit{template} module}
\label{fig:ast_example}
\end{figure}

\subsection{Tokenization of Names and Bodies}
This step converts the Ansible task descriptions (raw data units) to a stream of tokens, consumed by our deep learning algorithms. The names of the tasks are generally short texts in natural language, and thus we tokenize them by splitting them into words. However, the body of a task has a structured
representation. Hence, we use the abstract syntax tree (AST) of the task body to generate the token sequences while preserving the code's semantic information. In the research literature, ASTs are commonly used for representing code snippets as distributed vectors~\cite{SPOT, CODE2VEC}.
A task body defines an Ansible module's configuration and instance as a set of parameters (name-value pairs). It can also specify a conditional (\textit{when}, loop, and notify action to inform other tasks and handlers about the changes to the state of a resource managed by a module). We create an AST model that can capture these key information of a task body. To generate the token sequence from the AST, we use the pre-order depth-first traversal algorithm.
\Cref{fig:ast_example} shows a snippet of the generate AST model for the task example in \cref{fig:iac_snippet}. AST node types capture the semantic information such as modules and their parameters and notify action, and the raw code tokens capture the raw text values. The token stream generated from the AST will be \textit{[AnsibleTaskBody, Module, Name, template, Parameter, src, datadog.yaml.j2, ...., Notify, restart datadog-agent]}

\subsection{Generating Training, Test, and Validation Data}
Our linguistic anti-pattern detection is a binary classification task and employs supervised learning. Thus, we need a data set that includes correct (name-body consistent) and buggy (name-body inconsistent) task examples. As the Ansible is a domain-specific language and is relatively new, it is non-trivial to collect a sufficient number of buggy examples from real-world corpus. By inspired by the training data generation in the defect prediction literature~\cite{DeepBug, 9054826SANER}, we generate the buggy task examples from a given corpus of likely correct task examples by applying simple code transformations.
Before applying code transformations, we divide the tokenized data set into training, test, and validation sets to avoid potential data leakage between three sets during transformations. Within each data set, we swap the body of a given task with another randomly selected task to create inconsistencies. We consider two cases: (i) the tasks using the same module (\eg two tasks with the \textit{template} module) and (ii) the tasks using different modules (\eg one task with the \textit{template} module and another with the \textit{file} module). 

\subsection{Creating Vector Representations}
To feed the token sequences into a learning algorithm, we need to transform them into vector representations. We use the word embedding techniques for the vector representation of the Ansible task names and the corresponding task bodies. 
Word embedding techniques take a set of token sequences as inputs and produce a map between string tokens and numerical vectors~\cite{mikolov2013distributed}. They embed tokens into numerical vectors and place semantically similar words in adjacent locations in the vector space. As a result, the semantic information from the input text is preserved in the corresponding vector representation. We use \textsc{Word2Vec} to produce word embeddings. \textsc{Word2Vec} is a two-layer neural network that processes text by creating vector representations from words~\cite{mikolov2013distributed}. The input for the \textsc{Word2Vec} is a sequence of words (tokens), while its output is a set of feature vectors that represent these words. \textsc{Word2Vec} is used by several deep learning-based approaches to software defect prediction~\cite{DeepBug, SPOT,9054826SANER, DeepLiguistic}. 
We used the Continuous Bag of Words (CBOW) model to predict target words from the surrounding context words. The rationale is that IaC scripts (\ie task names and bodies) are sequences of tokens. Let us assume to have a sequence of tokens $t_1$,... $t_i$ ..., $t_j$) where $t_i$ is a token of an task, \textsc{DeepIaC} considers a window of $w$ tokens around $t_i$. For predicting the context of the token $t_i$, the two methods consider $\frac{w}{2}$ tokens before $t_i$ and $\frac{w}{2}$ tokens after $t_i$. 

\subsection{Training Predication Models}
We use Convolutional Neural Network (CNN) to build our binary classifier that can categorize the tasks into name-body consistent or not. CNNs are biologically-inspired variants of multi-layer artificial neural networks~\cite{matsugu2003subject}. Although they are widely used in image classification tasks, numerous studies report their success in the domain of NLP~\cite{kim2014convolutional, wang2015semantic}, and defect prediction of textual source code~\cite{DeepBug, allamanis2016convolutional, SPOT, DeepLiguistic}.

\begin{figure}
\centering
\includegraphics[width=1\columnwidth]{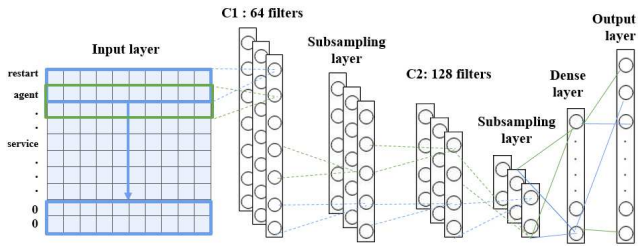}
\caption{Architecture of CNN model used}
\label{fig:cnn_design}
\end{figure}

\Cref{fig:cnn_design} shows the architecture of CNNs that our approach uses. The embedded token vectors of Ansible tasks generated by the trained CBOW (\textsc{Word2Vec}) model are used as an input to a CNN. The CNN input is a two-dimensional vector whose dimension varies since each Ansible module has a different vocabulary. However, each vector representation is long 100 words since we use the same setup for training \textsc{Word2Vec}. We use two convolutional pooling layers to reduce data dimensionality and capture the tasks' local features, similarly to the previous work~\cite{SPOT}.
L2 regularization is used in each convolutional layer. Furthermore, a dropout layer avoids overfitting, and a dense layer combines the previously captured local features by the convolutional and subsampling layers. The dense layer's output vector predicts and detects inconsistent module use within a task. The output layer consists of neurons per one task of each module. The output neurons are 0 (inconsistent) or 1 (consistent). The measure for the loss function is the \textsc{Mean Absolute Error} (MAE), and the corresponding optimizer is the \textsc{Stochastic Gradient Descent} (SGD).
We padded the input token sequences for the CNN training to comply with the fixed-width input layer on CNN. Motivated by Wang \etal~\cite{wang2016automatically}, we appended zero vectors at the end of the token sequences to reach the size of the longest token sequence of the input tasks. To compute the maximum length of the input sequences $s$ we used the equation: $max\_length_s = mean_s + standard\_deviation_s$.
To avoid having long sequences with many padded zeros, we decided the max length of the input sequences should be within two standard deviations of the mean \cite{moore2015basic}. This way, we filtered outliers by reducing noise from the padded zeros, and only the $3\%$ of the input token sequences were affected by this operation.

\subsection{Inconsistency Identification}
Inconsistency identification is a binary classification task since the test data are labeled in two classes: \textit{consistent} (the negative class in this work) and \textit{inconsistent} (the positive class in this work).
Once the binary classifier is trained with a sufficiently large amount of training data, we can query it to predict whether unseen Ansible tasks (\eg unseen test data sets) have name-body inconsistencies. 
To evaluate the performance of the trained models, we used the common metrics used in binary classification problems, namely \textit{accuracy}, \textit{precision}, \textit{recall}, \textit{F1 score}, \textit{MCC} (Matthews correlation coefficient), and \textit{AUC} (Area Under the \textit{ROC} (receiver operating characteristic curve) Curve).

\subsection{Implementation} 
To parse Ansible tasks and build ASTs for them, we developed a custom python tool. We tokenized the task names using the NLTK library\footnote{\url{http://www.nltk.org/}}. We used the Word2vec implementation of the gensim library to generate vectors from tokens. We implemented CNN/deep learning models using TensorFlow and Keras frameworks. We used PyGithub\footnote{\url{github.com/PyGithub/PyGithub}} and PyDriller~\cite{spadini2018pydriller} to locate repositories that contain Ansible IaC scripts. The complete prototype implementation of DeepIaC, including data set is available on GitHub\footnote{\url{github.com/SODALITE-EU/defect-prediction/tree/master/ansible}}.

\section{Empirical Evaluation}\label{sec:evaluation}
We evaluate DeepLaC by applying it to a real-word corpus of Ansible tasks. We aim to answer the research question: \textit{How accurately can we detect linguistic anti-patterns in infrastructure as code (IaC) using a Deep-Learning approach?}

\subsection{Data Collection}
We collected the data set from GitHub. To ensure the quality of the data collected, we used the following criteria (adopted from Rahman \etal \cite{rahman2018characterizing}) when searching for repositories that include Ansible scripts.

\begin{description}[leftmargin=*]
\item[Criteria 1] At least 11\% of the files belonging to the repository must be IaC scripts.
\item[Criteria 2] The repository has at least 10 contributors.
\item[Criteria 3] The repository must have at least two commits per month.
\item[Criteria 4] The repository is not a clone.
\end{description}

We found 38 GitHub repositories that meet the above criteria. We extracted $18,286$ Ansible tasks from them. As we trained a CNN model per a unique Ansible module, our experiments only considered 10 most used modules, which account for $10,396$ tasks in the collected data set. As discussed in \Cref{sec:approach}, we split each task into its name and body and tokenized both.

\subsection{Data Preparation and Model Tuning}

We split the tokenized dataset as follows: 60\% of the data was used for training, 20\% was the test set during the training, and 20\% was used to evaluate our model. We applied the transformations described in Section 3.3 to create the corresponding buggy data sets. The filtered token sequences were the input to the Word2vec. We tuned the Word2Vec parameters as: model(CBOW), vector size(100), Learning rate (0.025), Min word frequency(1), window size(6), and epochs(1000).
Next, we embedded the vector representations to the CNN classifier. We tuned the parameters of the CNN as convolution dimension (10), activation layer (ReLu), output layer (softmax), optimization algorithm (sgd), token sequence range (84-99), learning rate (1e-02
), pooling type (max pool), and loss function (MAE).

\subsection{Effectiveness in Identifying Inconsistencies}

\begin{table*}[ht]
\centering
\caption{Classification results for the top 10 used Ansible modules}
\begin{adjustbox}{max width=\textwidth}
\begin{tabular}{|c|r|c|c|c|c|c|c|c|c|c|c|}
\hline
\multicolumn{2}{|l|}{\textbf{Evaluation Metric/Module}} & \textbf{shell} & \textbf{command} & \textbf{set\_fact} & \textbf{template} & \textbf{file} & \textbf{gather\_facts} & \textbf{copy} & \textbf{service} & \textbf{debug} & \textbf{fail} \\ \hline
\multirow{3}{*}{\textbf{Inconsistent}} & Precision & 0.880 & 0.790 & 0.770 & 0.820 & 0.900 & 0.900 & 0.860 & 0.870 & 0.870 & 0.820 \\ \cline{2-12} 
 & Recall & 0.810 & 0.840 & 0.900 & 0.940 & 0.940 & 0.830 & 0.810 & 0.760 & 0.770 & 0.690 \\ \cline{2-12} 
 & F1 score & 0.843 & 0.814 & 0.830 & 0.876 & 0.920 & 0.864 & 0.834 & 0.811 & 0.817 & 0.749 \\ \hline
\multirow{3}{*}{\textbf{Consistent}} & Precision & 0.810 & 0.820 & 0.890 & 0.930 & 0.930 & 0.905 & 0.82 & 0.800 & 0.750 & 0.760 \\ \cline{2-12} 
 & Recall & 0.890 & 0.770 & 0.750 & 0.800 & 0.890 & 0.770 & 0.870 & 0.900 & 0.860 & 0.870 \\ \cline{2-12} 
 & F1 score & 0.848 & 0.794 & 0.814 & 0.860 & 0.910 & 0.870 & 0.844 & 0.847 & 0.801 & 0.811 \\ \hline
\multicolumn{2}{|r|}{Accuracy } & 0.847 & 0.805 & 0.819 & 0.868 & 0.915 & 0.817 & 0.838 & 0.833 & 0.809 & 0.785 \\ \hline
\multicolumn{2}{|r|}{MCC} & 0.697 & 0.610 & 0.649 & 0.744 & 0.830 & 0.685 & 0.678 & 0.669 & 0.625 & 0.570 \\ \hline
\multicolumn{2}{|r|}{AUC} & 0.848 & 0.804 & 0.822 & 0.868 & 0.914 & 0.848 & 0.838 & 0.830 & 0.814 & 0.779 \\ \hline
\end{tabular}
\label{tab:word2vec_part1}
\end{adjustbox}
\end{table*}

\Cref{tab:word2vec_part1} presents the inconsistency detection results for the top 10 Ansible modules in our data set. Overall, our approach yielded an accuracy ranging from $0.785$ to $0.915$, AUC metric from $0.779$ to $0.914$, and MCC metric from $0.570$ to $0.830$. Our approach achieved the highest performance for detecting inconsistency in the \textit{file} module, where the accuracy was $0.915$, the F1 score for the inconsistent class was $0.92$, and the F1 score for the consistent class was $0.91$. We also observed that the ROC curve, the model loss, and the accuracy plots confirm the model's good performance. Due to the limited space, we do not present the corresponding visualizations; however, they are available in the GitHub repository of this study.

\subsection{Threats to Validity}
\paragraph{Threats to construct validity} The collected repositories may not be relevant for the problem at hand. 
We mitigated this threat by applying the criteria used in previous works on IaC to ensure the collected data's quality.
Although the number of repositories may seem low, a small but relevant and representative dataset of active repositories is preferable.
Another threat to construct validity concerns the mutation of scripts employed to generate inconsistent cases, which may not represent real-world bugs.
Nevertheless, we tried to mitigate this threat by applying the existing approaches that have successfully used mutation to generate the training data~\cite{DeepBug,9054826SANER}. 
We plan to further mitigate this threat by gathering more real-cases of inconsistent tasks.

\paragraph{Threats to internal validity} The choice of the features used to train the CNN model could influence linguistic anti-patterns detection.
We mitigated this threat by training the model using a high number of features (obtained by transforming each task to a vector space of words) extracted from more than ten thousand Ansible tasks.
The feature engineering for the classification task depends on the quality of the code base, including naming conventions, typos, and abbreviations. This aspect poses a threat to validity, and advanced NLP techniques can be employed to overcome this. 

\paragraph{Threats to external validity} The conclusions are derived only from a subset of modules in Ansible (\ie the ten most used), which might not be reproducible for other modules and languages.
However, we used both generic modules (such as \textit{command} modules) and more specific modules. Specific modules (\eg the \textit{copy} module) do focus works, but general modules can execute ad-hoc OS commands. 
We believe that using a mix of generic and specific modules may mitigate, at least partially, this threat.
Finally, we analyzed only Ansible projects, and the results could not generalize to other IaC languages (\eg Chef, Puppet).
Extend our approach to such languages is part of our agenda.

\section{Conclusion and Future Work}\label{sec:conclusion}
\textsc{DeepIaC} detects linguistic anti-patterns in IaC scripts by leveraging word embedding and deep learning. In particular, it provides automated support to the users to debug inconsistencies in the names and bodies of IaC code units. Our experimental results show that our approach's performance achieves an accuracy between $0.785$ and $0.915$ in detecting inconsistencies. We plan to extend the \textsc{DeepIaC} to detect name-based bugs~\cite{DeepBug} and misconfigurations in IaC code scripts. We also aim to apply it to other IaC languages.

\begin{acks}
This work is supported by the European Commission grant no. 825480 (SODALITE H2020) and no. 825040 (RADON H2020).
\end{acks}

\balance

\bibliographystyle{ACM-Reference-Format}
\bibliography{main}

\end{document}